# Reversible State Vector Parallel Processor Modeled After Quantum Computer Behavior

John Robert Burger, Department of Electrical and Computer Engineering, California State University Northridge, Northridge, California

*Abstract* – Proposed below is a reversible digital computer modeled after the natural behavior of a quantum system. Using approaches usually reserved for idealized quantum computers, the Reversible State Vector Parallel (RSVP) processor can easily find keywords in an unstructured database (that is, it can solve a needle in a haystack problem). The reversible processor efficiently solves a SAT (Satisfiability of Boolean Formulae) problem; also it can aid in the solution of a GP (Global Properties of Truth Table) problem. The power delay product of the reversible processor is exponentially lower than that of a standard CAM programmed to perform similar operations.

**Introduction**

As a phenomenon of nature, quantum computation is mainly a subject of natural science [1, 2]. In contrast, computer science is largely a science of the artificial [3]. Occasionally these strongholds communicate. For example, it is fairly common, and occasionally useful to digitally simulate quantum computers [4, 5].

The author now moves beyond simulators to discover from nature new ways to accomplish reversible, parallel computing. Much thought has been given to reversible gates, especially the unconditional NOT (or UCN), the Single Controlled NOT (or SCN) and the Double Controlled NOT (or DCN) along with proposed applications [6, 7]. Computers that run either forward or reverse, like an automobile, or a motion picture, have been discussed extensively [8, 9].  However little is advertised about practical architectures for large-scale reversible computers.  Presented below is a reversible state vector processor that is compatible with modern memory technology.  The logic gates themselves do not need to be reversible, although they could be.  Consider now the analogies in Table 1.

**Table 1. Quantum terms and computer science analogies**

|   | Quantum terms | Computer analogies |
|---|---|---|
| 1 | $|0>$, $|1>$ spin up, down; Basis vectors | Boolean 0, 1 |
| 2 | $|a> = z_1 |0> + z_2 |1>$ general spinors $|a> = (a_1 \ a_2)$ | None |
| 3 | $\mathbf{s} = |a>|b>|c> \ldots |n>$ state vector of n spinors in a system with $2^n$ states | $2^n$ addresses in a memory capable of holding a complex number. |
| 4 | $\mathbf{U} |a>$ unitary transformation such that $\mathbf{U} \ \tilde{\mathbf{U}}^* = \mathbf{1}$ | Reversible logic |
| 5 | observation forces $|a>$ to either $|0>$, $|1>$ | None |

Item 1 introduces the two-dimensional vectors $|0>$ and $|1>$.  They are termed qubits, an obvious deferment to Boolean bits 0 and 1.  Qubits obey the rules of quantum mechanics, but their physical form depends heavily on implementation.  Unlike ordinary 0 and 1, qubit vectors combine linearly (item 2) to form a" spinor."  To agree with physics observations, the coefficients $z_1$ and $z_2$ are complex numbers such that $|z_1|^2 + |z_2|^2 = 1$. $|z_1|^2$ and $|z_2|^2$ are the probabilities of observing either $|0>$ or $|1>$ when observation is attempted.

Item 3 suggests that n spinors can combine to form a state vector. A state vector appropriately formed with n qubits will have $2^n$ elements, each possibly a complex number. The elements of a state vector can be related to the elements of the component spinors. For example, one element would be the product $a_1 b_1 c_1 \ldots n_1$.

Item 4 suggests that state vectors can be transformed using a unitary transform. The reversible logic that can occur in a quantum computer is the result of unitary transformations. As noted in item 5, observation changes the state vector, destroying much of its information. Observation in quantum mechanics sets each $|a>$ to the result of the observation, either $|0>$ or $|1>$. The probabilistic nature a state vector, and the destructiveness of observation are a major challenge to quantum computation.

State vector representation in a digital computer

A state vector can be represented digitally. In Item 3 above, the state vector **s** for n qubits will have $2^n$ state values, where each is possibly complex, that is, an analog number with magnitude and phase. Obviously there would be truncation errors for non-integer analog values, because of finite register size. However, if the state values can be arranged to be integers, their representation in a digital computer can be exact.

The number of states is equal to $2^n$. Each state has an implied address. In a digital computer each address can be attached physically to each state value. Consequently, each artificial state contains not only a register for the state value, but also a register for the state value address.

Processing a state vector

The state vector, item 3 in the above table, has n qubits, and $2^n$ state values. In the natural world, computational leverage comes about when n qubits control $2^n$ states. Physically, qubits, for example, $|Num1\rangle|Num2\rangle$ take on linear combinations of basis states $|00\rangle, |01\rangle, |10\rangle$ and $|11\rangle$, each multiplied by some analog, possibly complex coefficient, appropriately normalized. These coefficients are defined to be the state values. State values are associated with state addresses, that is, basis state codes, for example, 00,01,10,11. Therefore, to execute digitally a unitary transformation that interchanges states, the state addresses are going to be modified in place, that is, stationary to the state values.

A useful tool, exemplified in Figure 1, is the address diagram.

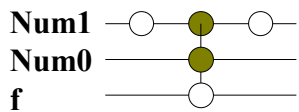

**Figure 1. Address Diagram.**

Assume Num1, Num0 are the bits of an address field; f is a flag bit. Bits are transformed according to symbols. The symbols, shown from left to right, are:

a) Unconditional NOT (UCN) = Empty circle for the bit that is acted upon. The acted upon bit is complemented.

b) Double Controlled NOT (DCN) = Vertical line with shaded circles for the two controls and an empty circle for the bit that is acted upon. If the controls are both true, then the acted upon bit is complemented.

Not shown are NOTs with only one control (SCN), and NOTs with multiple controls (MCN), readily available in the system presented below. The sequence of computations in an address diagram is reversible. If outputs are applied on the right, the original inputs result on the left.

An example of address processing is given in Table 2 with possible values of Num1, Num0. Three registers are assumed to hold the values of Num1, Num0 and an associated flag bit. Assume f is initially zero. Consider the case of Num1, Num0 = 0, 1. Inputting 010 (from top to bottom in the address diagram on the left), the unconditional NOT for Num1 in Figure 1 changes this to 110, the double-controlled NOT gate changes the code to 111, and the right most unconditional NOT restores Num1, Num0 back to 0, 1. The three registers are processed in parallel, but only the address 01 results in setting f.

**Table 2. Address scrambler result**

| Num1 | Num0 | f (Initial) | f (Final) |
|------|------|-------------|-----------|
| 0    | 1    | 0           | 1         |
| 0    | 0    | 0           | 0         |
| 1    | 0    | 0           | 0         |

Effectively, the code Num1, Num0 = 0, 1 was flagged.  This method could be applied to a large unstructured database to locate keywords fast.  Other uses for reversible computation are mentioned later.

**Architecture To Process An Address Diagram**

When a quantum mechanical vector undergoes a given unitary transformation, it is convenient to imagine that state values undergo changes without significant time delay, so that the changes can be modeled as occurring in parallel.  This suggests using concurrent processors as in Figure 2.  Each address register, for example, $WORD_0$ is initialized to a unique address.  Associated with each address register is a data register, for example, $D_0$ that represents digitally the value of the state.  Although state values could be made available, usually they are unnecessary.  In all applications mentioned below, the data register is empty.

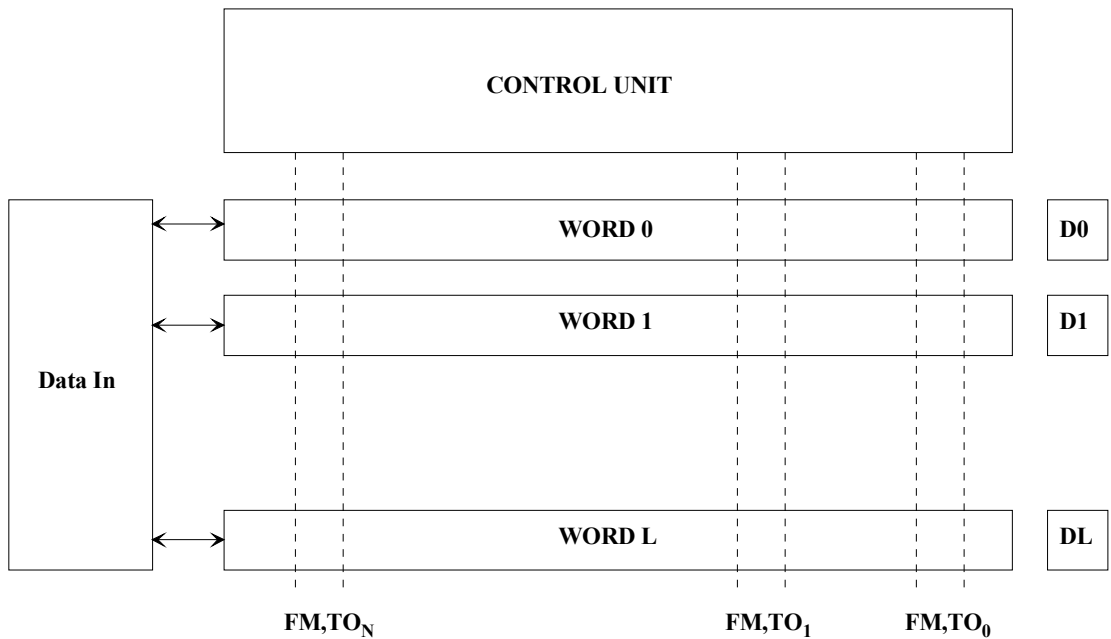

**Figure 2. Reversible State Vector Parallel (RSVP) Processor For L words, n+1 Bits**

What must be accomplished in address processing is very simple: If controlling bits in a particular word are true, than a destination bit in that word must be complemented. Electrically, such operations are particularly easy to implement using the concept of a T flip-flop. Figure 3 shows a word design. From the control unit there are two wires for each cell. For example, $FM_o$, $TO_o$ go to all of the $A_0$ in each word. TO is a signal that activates the subject (destination) bit. FM is a signal that selects individual controlling (source) bits within each register. Selected source bits in turn enable the bus labeled FMbus. The logic is such that if the source bits (doing the controlling) are active, the destination bit or bits will be complemented.

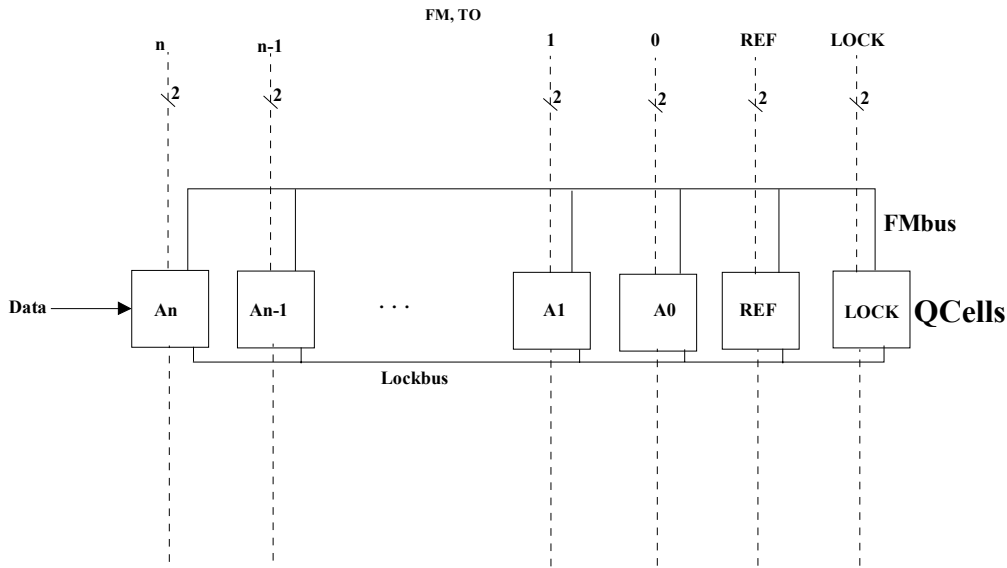

**Figure 3. Word structure showing REF and LOCK cells.**

 The purpose of the REF cell is to provide an unconditional true for implementing unconditional NOTs. In fact, any number of such NOTs can be executed within any one step of an address diagram (where ever TO signals are true). The purpose of the cell labeled LOCK is to lock out unwanted usage of the bus, to save power.

Each cell is initialized in some convenient way using the Data input. It is useful to define a given bit, for example $A_0$ to be a flag bit. If a flag goes true in an assembly of such registers, it is usually desired to know exactly where the flag is located. A reasonable approach is to shift out flags until a true one is found. Alternately, there are other possibilities using build-in decoder logic to locate a true flag. Generally keywords would not be modified and would not be read out (unless there was a compelling reason to do so).

**Performance Analysis – RSVP vs. CAM**

Q cells are defined to be memory cells that store and process a single bit, for example, $A_o$. Figure 4 shows Q Cell logic (not suggested as an implementation). The three-state buffer will be open (its control will be low). FMbus will be pulled high if the data in the T flip-flop, that is, D is true, while LOCK is low (LOCK-bar is high) and FM is high. If D is false under these conditions, the three-state buffer will drive the bus low. If TO is true and the bus is true, the flip-flop will toggle as required in a RSVP processor.

Specialists in memory cells will recognize the Q cell as related to a CAM cell (content addressable memory) [10], yet there are major differences. The Q cell is based on a toggle or T flip-flop, whereas CAM cell is based on a D flip-flop.

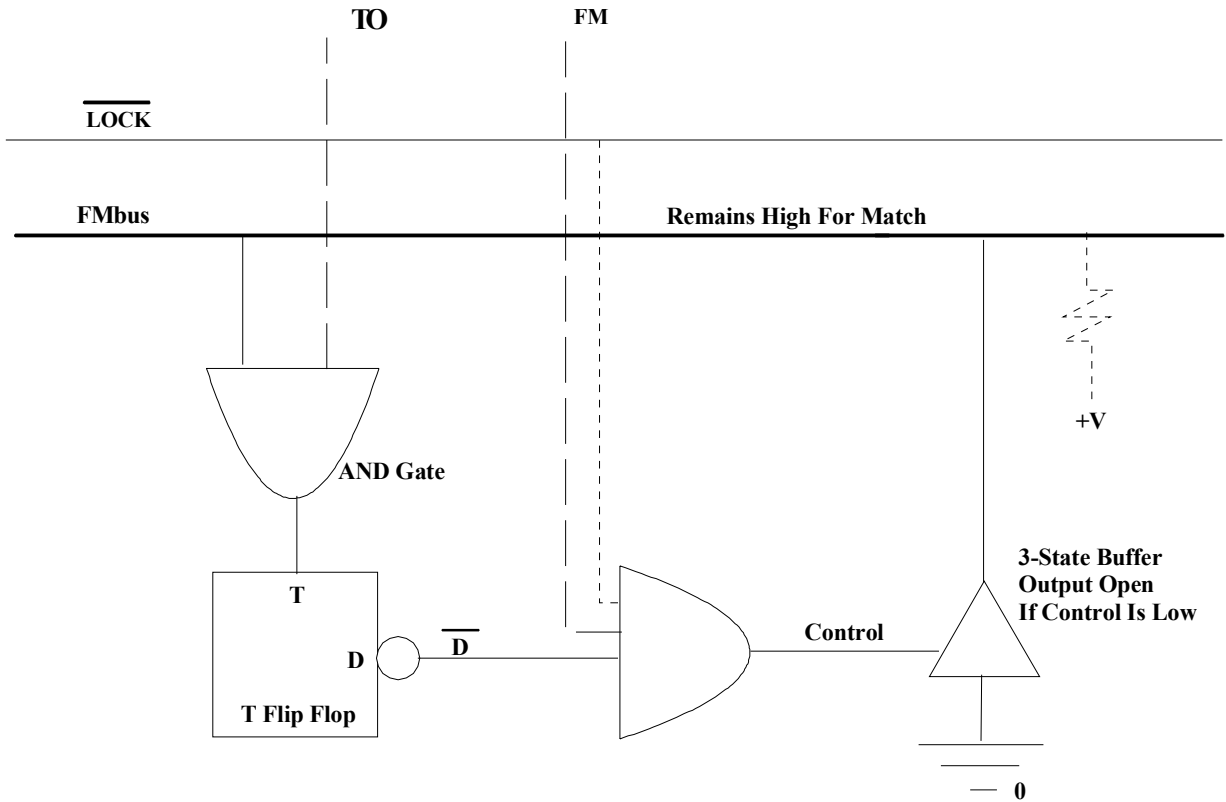

**Figure 4. Q Cell Circuit. (This is <u>not</u> a suggested implementation).**

Figure 5 shows a generic CAM system. CAM data must be read-able and write-able, so there must be multiple match resolution and multi read logic as well as multi write capability. An outline of a procedure to implement a controlled NOT in this CAM has the following parts:

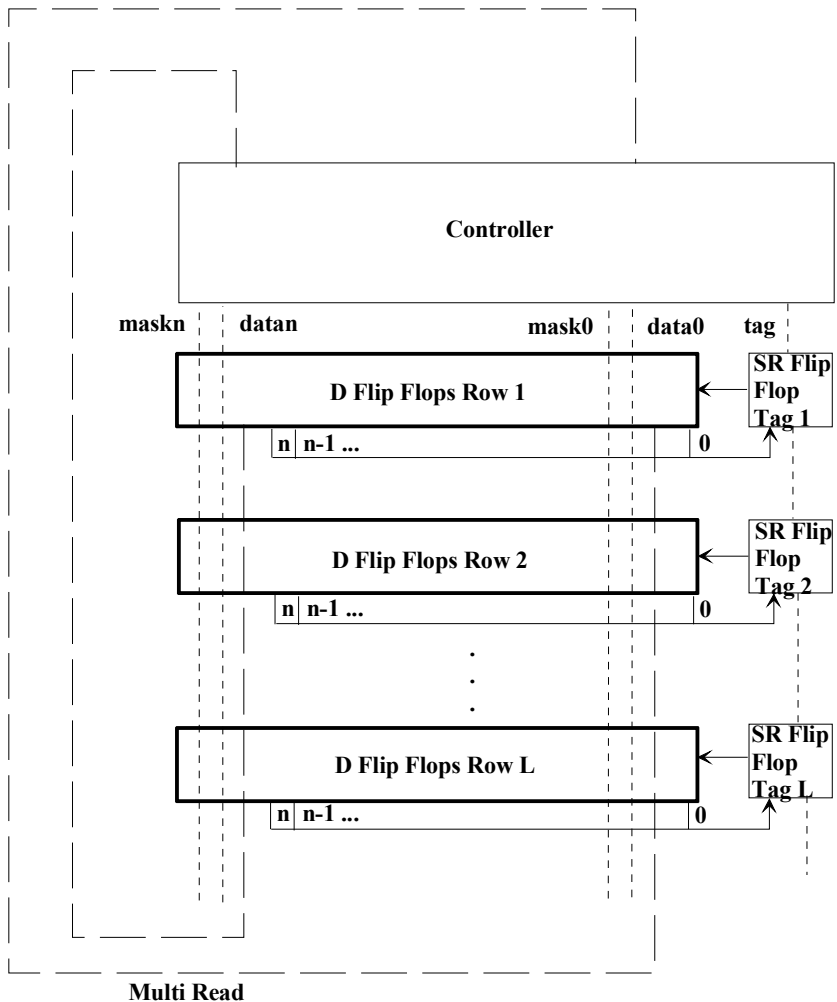

**Figure 5. CAM System**

Step 1. Reset all the tags and all mask bits; set the mask bits for those bits that are being tested; set the data buses Boolean one. Mask, tag and data signals travel from the controller to the rows of the CAM.

Step 2. Enable a search for sought after true bits. Signals travel from rows to the tag flip-flops.

Step 3. Read the rows for which the tag is true beginning with the topmost row. Complement the data. Write it back to the row it came from and reset that row's tag.

Note that signals travel from the row to the controller, are processed, and then travel from the controller to the row again. This is done for each true tag.

To estimate delays and power dissipations, it is assumed that each bus line is active and is initialized halfway between true and false. Subsequently each bus goes slightly high or slightly low to activate a regenerative sense amplifier. The amplifier pushes the bus to its valid logic. Power dissipation at the location of the flip-flops is minimal.

A comparable procedure for a Reversible State Vector Parallel Processor is as follows:
Step 1: Control unit (Figure 2) sets appropriate FM and TO signals; signals travel to the words.
Step 2: Words (Figure 3) execute bit reversals asynchronously as signals travel across the word. There are no tags and there are no multi reads in the author's concept.

**Speed**

Each cycle involves sending signals across the chip, and can be made to take comparable amounts of time. The CAM takes a unit of time $\Delta_1$ for step 1; a unit of time $\Delta_2$ for step 2; and $2\Delta_1$ for each cycle of step 3. The CAM is assumed to work it way through a long series of multi reads and multi writes because a large number of matches are expected. For n+1 bits in a row, the number of binary counts could be $2^{n+1}$. In a binary count, ½ of all counts will be matches for SCN; ¼ will be matches for DCN and so on. Let f be the fraction of the total that are matches. The number of matches will be $f\,2^{n+1}$. The net time for a CAM operation is about $\Delta_1 + \Delta_2 + 2\,f\,\Delta_1 2^{n+1}$. In contrast, the RSVP processor uses about $\Delta_1 + \Delta_2$ (or less since a tag does to have to be set) so in general, CAM needs exponentially more time to operate, assuming multiple matches as expected in state vector processing.

**Power**

The total number of signals being sent across the chip in a CAM basically affects power dissipation for a given operation.

<u>Step 1</u> – Signals go from the controller to the rows, where each vertical line is assumed to take a power of $P_1$. The total for this is about $P_1(n+2) \approx P_1 n$.

<u>Step 2</u> – Signals, L in number, go from the rows to the tags, where each horizontal line is assumed to require a power dissipation of $P_2$. The total is about $P_2 L$

<u>Step 3</u> – Signals go from the rows to the controller, and back to the rows, using about $2P_1 n$ per cycle. Unfortunately, cycle 3 is repeated $f\,e^{n+1}$ times, assuming multiple matches. The net result to execute a controlled NOT is that power is approximately $P_1 n + P_2 L + 2\,f\,P_1\,n\,e^{n+1}$.

In the case of a Reversible State Vector Processor, each cycle involves TO and FM signals for each of the n+1 Cells. These vertical buses are assumed to take about $P_1 n$. Then the horizontal buses take about $P_2 L$ (or less since a tag does to have to be set). The main difference is Step 3, which is expected to be exponentially large. CAM uses exponentially more power and exponentially more delay. For reversible computations on a large scale, a CAM system would need to be designed and fabricated to imitate RSVP, and to be RSVP processing.

**Performance During Reversible Computation**

Power dissipation in a Reversible State Vector Parallel Processor is actually lower than what the above discussion indicates because in a sequence of address operations, many words can be disabled from the power equation. This is done with the lockbus shown in Figure 3. For example, assume a search for x,y = 1, 1. In a normal binary count only ¼ of the words will have both x and y true. Subsequently ¾ of the words can have their horizontal buses can be disabled. The associated reduction in power dissipation can be important in a long series of operations in which given words are known to <u>not</u> contain the sought after information. Applications are discussed further below.

1. Keyword Problem – The goal is to locate a keyword in a large unstructured database. How it works can be explained with a simple example. Imagine that the code in Table 3 represents three unstructured keywords. What the keywords point to is assumed stored in mass storage, separate from keyword memory, and in the same order as they appear in keywords memory.

**Table 3. Keyword Initialization**

| Num1 | Num0 |
|------|------|
| 0    | 1    |
| 0    | 0    |
| 1    | 0    |

Keywords have an extra bit labeled f that is initialized to zero as in Table 4. This will serve as a flag bit.

**Table 4. Flag Bit Initialization**

| Num1 | Num0 | f |
|------|------|---|
| 0 | 1 | 0 |
| 0 | 0 | 0 |
| 1 | 0 | 0 |

Keyword search proceeds by applying the address sequencer as in Figure 2 above. The sought after keyword is immediately identified with a flag that goes true. This application was inspired by Grover's Needle-in-the-haystack algorithm [11].

Advantages of the above system are: (1) Search is done in parallel and is faster than a serial method. (2) Data may be unordered. Thus it is possible to search for keywords that are neither alphabetical nor numerical. (3) It is convenient to have far fewer items than some power of two.

If random access memory has unordered information, a brute force search would have to search through (or hash) an exponentially large number of items, taking exponential time. Time in a RSVP processor is proportional to the number of address bits being searched, so search time is linear (and faster).

2. <u>SAT (Satisfiability of Boolean Formulae) Problem</u> – Sometimes solutions are hidden, although a Boolean function can be constructed that knows a solution when it sees one. Any binary function can be expressed in an address diagram, so solutions to SAT problems are available, as long as the numbers not too large. As an example involving simple arithmetic, research is proceeding on algorithms to find concurrently <u>all</u> divisors of a composite number [12]. As an example involving graphics, research is proceeding to find <u>all</u> Hamiltonian cycles in a graph [13]. These approaches were inspired by investigations into the power of a quantum computer [1]. A RSVP computer can solve a SAT problem in linear time. Ordinary classical computers are inefficient for SAT problems because they have to go through an exponential number of trials.

3. <u>GP (Global Properties of Truth Table) Problem</u> – The truth table of a binary function has global properties that are of interest in cryptography. Examples are: a balanced number of 1s and 0s in its truth table, symmetry in the truth table, and periodicities in the truth table. Note there are truly a large number of binary functions for a given number of bits n. There are $2^{**}(2^{**}n)$, where $2^{**}n = 2^n$, so finding the global properties of a function with a large n is nontrivial.

A RSVP processor can help identify global properties [14]. For example, assume the goal is to determine the global properties of an unknown Boolean function whose truth table is delineated as in Table 5.

**Table 5. Global Properties**

| Num1 | Num0 | f |
|------|------|---|
| 0 | 0 | 0 |
| 0 | 1 | 1 |
| 1 | 0 | 1 |
| 1 | 1 | 0 |

It is clear to a trained eye that f has multiple levels of anti symmetry (in a block whose size is a power of two, each corresponding term in the bottom half is the opposite of the respective term in the top half). Such global properties can be expressed as a code. In the above example the code could be 11. This means that f has anti symmetry, and that f has sub blocks with anti symmetry.

Classical evaluation of binary functions with a large number of bits requires exponential work, and takes too long, because the length of a truth table grows as a power of two. However, since the number of steps to evaluate a function in the above system is proportional to number of bits, it is relatively easy to work with a large truth table. By reading out the flag bits, it is possible to identify global properties in a post-processor. Thus a GP Problem is solvable, whereas otherwise such problems are intractable in a classical computer for large numbers of bits.

**Conclusion**

By imitating the way a quantum system works, a plan for a Reversible State Vector Parallel (RSVP) Processor is given above that will support reversible operations according to an address diagram. The addresses of each state are held in conjunction with the state itself in a special register. There will be a large array of such registers each initialized differently, but each receiving similar instructions in parallel from an address diagram. The processing amounts to complementing a given bit if selected other bits in the address are true. This interchanges the addresses of states, which if done cleverly, yields a useful result.

Obvious above is that a RSVP processor takes not only lower, but exponentially lower power-delay product than a CAM doing similar operations. This is mainly due to the assumption of multi reads and writes in order to complement bits.

References are made above to unstructured keyword search (Needle in a haystack), SAT (Satisfiability of Boolean Formulae), and also GP (Global Properties of Truth Table) problems. A reversible computer of the type outlined above, if properly built, accomplishes such tasks in linear time, whereas a conventional computer takes an impractical exponential length of time.

*Acknowledgements* – The author thanks Dr. Won Woo Ro and Dr. Robert Henderson for their helpful suggestions.